# Supersonic motion of high frequency vibrational soliton in monatomic chain

V. Hizhnyakov

Institute of Physics, University of Tartu, W. Ostwaldi Str 1, 50411, Tartu, Estonia

**Abstract.**

Moving self-localized vibration of large size (vibrational soliton) in monatomic chain with cubic and quartic anharmonicities is considered. Two types of motion of such vibrations are found: the subsonic and the supersonic ones. First type of motion is possible for chains with hard quartic anharmonicity supposing that cubic anharmonicity is weak. This motion was earlier observed in numerical simulations. For the second type of motion the cubic anharmonicity plays the decisive role: the moving mode is stabilized by co-moving deformation of the lattice caused by this anharmonicity. The quartic anharmonicity in this case may be absent or even soft.

**Introduction**

Existence of spatially localized stable vibrational excitations in perfect atomic lattices is one of the basic effects of nonlinearity of matter. One type of such excitations is intrinsic localized modes (ILMs) (called also discrete breathers (DBs)) [1-8]; these modes appear due to anharmonicity of the lattices and they have frequencies outside of the phonon spectrum. Unlikely to usual solitons in continuous media, ILMs (DBs) are spatially localized. The simplest modes of this type are the self-localized anharmonic modes in monatomic chain considered by Kosevich and Kovalev [1]. These modes have large size (as compared to interatomic distance) and frequency slightly higher than the upper limit of the phonon spectrum. They exist in chains with predominantly hard quartic anharmonicity.

In recent years it has been found that self-localized anharmonic modes (ILMs) can move and therefore they can be called vibrational solitons; see, e. g. Refs. [9-11], where mobility of ILMs was numerically demonstrated in atomic chains, and Ref. [12], where the mobile ILMs were numerically found in several metals. The mobility of ILMs (DBs) was also discussed in Refs. [13-15]. From presented results one can conclude that the reason of moving of large size ILMs can be, e.g. invariance of equations of motion under linear transformation of coordinate and time for the envelope wave. Note that this condition is fulfilled for the self-localized vibrations of Kosevich and Kovalev; thus one expects that they can move. Numerical simulations confirm this expectation: according to these simulations this mode indeed moves without any energy loss [9,10]. The observed velocity of the motion is small in comparison to the velocity of sound. Surprisingly, as far as we know, no analytical consideration of the motion of this simplest ILM has been presented so far. Here such a consideration is given.

Actually we considered envelope waves of self-localized anharmonic modes, both spatially localized and moving. The DC component (lattice distortion) of the modes is present in both cases; the latter appears due to cubic anharmonicity and it is stabilized by co-moving DC–component. We have found that besides slowly (as compared to sound velocity) moving ILMs known from the numerical simulations, in this model the fast - supersonic motion of ILMs is possible. For the latter motion the cubic anharmonicity plays a decisive role: it causes the concomitant deformation of the chain which stabilizes the motion. The quartic anharmonicity in this case may be absent or even soft. We also have found that the motion of the self-localized mode results in the increasing of its frequency and in the self-modulation causing decrease of the frequency in the first edge and increase of it in the trailing edge of the excitation wave.

**Equation of motion for envelope wave in monatomic chain**

Here we consider the large size self-localized anharmonic modes of Kosevich and Kovalev in monatomic chain described by the following potential energy:

$$W = \sum_n \left( \frac{1}{2}(u_n - u_{n-1})^2 + \frac{1}{3}\lambda(u_n - u_{n-1})^3 + \frac{1}{4}\mu(u_n - u_{n-1})^4 \right) \qquad (1)$$

Here $u_n$ is the longitudinal displacement of the atom number $n$, $\lambda$ and $\mu$ are the cubic and quartic anharmonicity constants (the units $K_2/M = 1$ are use, where $K_2$ is the elastic force constant, $M$ is the mass of the atoms). Following Ref. [1] we consider the envelope wave of this mode. It appears more convenient to regard odd and even displacements

$$\chi_n = u_n - u_{n-1}$$
$$\psi_n = u_n + u_{n-1} \qquad (2)$$

where $\chi$ is the difference in the displacements of two neighboring atoms, while $\psi$ defines twice the displacement of the center of gravity of the two neighboring atoms. The displacements $\chi_n$ account for the fast oscillating in time motion of the distance between atoms (with the frequency $\omega$ above, but close to the top phonon spectrum $\omega_M = 2$ and it corresponds to AC-component of the KK mode. Unlikely to that the displacements $\psi_n$ are almost time-independent and they correspond to the DC-component of the mode – the static distortion of the chain caused by the KK mode. The authors of Ref. [1] consider the case when the amplitudes of $\chi_n$ and the displacements $\psi_n$ slowly vary with coordinates of atoms. This allows them to restrict the consideration with the second derivatives of $\chi$ and $\psi$ with respect to the coordinate $x = a_0 n$. Obtained in this way system of equations for $\chi$ and $\psi$ reads [1]

$$\frac{\partial^2 \chi}{\partial t^2} + \frac{\partial^2 \chi}{\partial x^2} + 4\chi - 4\lambda\chi\frac{\partial \psi}{\partial x} + 4\mu\chi^3 = 0 \qquad (3)$$

$$\frac{\partial^2 \psi}{\partial t^2} - \frac{\partial^2 \psi}{\partial x^2} + 4\lambda\chi\frac{\partial \chi}{\partial x} = 0 \qquad (4)$$

**Spatially localized envelope wave [1]**

Let us regard first the immobile (spatially localized) mode. We consider that $\chi$ can be presented in the form $\chi = \bar{\chi}\cos(\omega t)$, where $\bar{\chi}$ does not depend on time; and $\psi$ does not depend on time either. (Existing higher-order harmonics with frequencies $2\omega, 3\omega,...$ for large size modes are very weak and can be neglected.) Therefore Eqs. (3) and (4) can be presented in the form

$$\left(-\omega^2 + 4\bar{\chi} + \partial^2/\partial x^2\right)\bar{\chi} - 4\lambda\bar{\chi}\,\partial\psi/\partial x + 3\mu\bar{\chi}^3 \cong 0 \qquad (5)$$

$$\partial^2\psi/\partial x^2 \cong \lambda\,\partial\bar{\chi}^2/\partial x \qquad (6)$$

Here we take into account relations $\cos^3(\omega t) = (3\cos(\omega t) + \cos(3\omega t))/4$ and $\cos^2(\omega t) = (1 + \cos(2\omega t))/2$ and exclude the terms $\propto \cos(3\omega t)$ and $\propto \cos(2\omega t)$ from the consideration. Integrating the last equation we get

$$\partial \psi / \partial x = \lambda \bar{\chi}^2 \tag{7}$$

As a result, Eq. (3) gets the form:

$$(-\omega^2 + 4 + \partial^2/\partial x^2)\bar{\chi} + (3\mu - 4\lambda^2)\bar{\chi}^3 \cong 0 \tag{8}$$

Solving this equation one gets [1]

$$\chi \cong \sqrt{\frac{2}{3\mu - 4\lambda^2}} \frac{\varepsilon \cos(\omega t)}{\cosh(\varepsilon x)}, \tag{9}$$

$$\psi = \frac{2\lambda\varepsilon}{3\mu - 4\lambda^2} \tanh(\varepsilon x), \tag{10}$$

where $\varepsilon = \sqrt{\omega^2 - \omega_M^2} \ll \omega$. This is the equation for the immobile self-localized anharmonic mode.

**Moving envelope wave**

Let us consider first moving mode in case of the chain with purely quartic anharmonicity. In this case the DC component of the mode is absent $\psi = 0$ and Eq. (3) gets the form

$$(\partial^2/\partial t^2 + \partial^2/\partial x^2)\chi + 4\chi + 4\mu\chi^3 = 0. \tag{11}$$

The variables $t$ and $x$ enter Eq. (11) in the same way. Therefore a rotation-type transformation of these variables

$$\begin{cases} t' = \cos(\varphi)t + \sin(\varphi)x \\ x' = \cos(\varphi)x - \sin(\varphi)t \end{cases}, \tag{12}$$

which corresponds to moving reference frame with velocity

$$v = \tan(\varphi), \tag{13}$$

leaves Eq. (11) unchanged:

$$\frac{\partial^2 \chi'}{\partial t'^2} + \frac{\partial^2 \chi'}{\partial x'^2} + 4\chi' + 4\mu\chi'^3 = 0.$$

The solution of this equation is $\chi' = \sqrt{2/3\mu} \cdot \varepsilon \cdot \cosh^{-1}(\varepsilon x') \cdot \cos(\omega t')$. In the laboratory reference this solution reads

$$\chi' = \sqrt{\frac{2}{3\mu}} \cdot \frac{\varepsilon \cos(\omega'(t + vx))}{\cosh(\varepsilon'(x - vt))}, \tag{14}$$

where $\omega' = \omega/\sqrt{1+v^2}$, $\varepsilon' = \varepsilon/\sqrt{1+v^2}$. Note that $v=1$ corresponds to the velocity of sound. The time-dependence of the excitation at the maximum corresponds to oscillations with the frequency $\omega\sqrt{1+v^2} > \omega$. This means that the motion of the ILM is accompanied by an increase of its frequency. Besides, at the forward edge of the moving ILM the frequency is enlarged while at the backward edge it is reduced. This is the result of self-phase modulation caused by the nonlinearity.

**Account of cubic anharmonicity**

Presented above derivation of the mobile envelope wave in the chain with quartic anharmonicity can be easily extended to the chains with cubic and quartic anharmonicity. To this end, we consider the solutions $\chi' = \chi'(x',t') \cong \cos(\omega t')\bar{\chi}'(x')$ and $\psi' = \psi'(x')$. We will show that equations

$$\frac{\partial^2 \chi'}{\partial t'^2} + \frac{\partial^2 \chi'}{\partial x'^2} + 4\chi' - 4\lambda\chi'\frac{\partial \psi'}{\partial x} + 4\mu\chi'^3 = 0, \quad (15)$$

$$\frac{\partial^2 \psi'}{\partial t^2} - \frac{\partial^2 \psi'}{\partial x^2} + 4\lambda\chi'\frac{\partial \chi'}{\partial x} = 0 \quad (16)$$

can be satisfied. Taking into account that $\psi'$ depends only on $x'$ we find

$$\partial^2 \psi'(x')/\partial t^2 = \tan^2(\varphi)\partial^2 \psi'(x')/\partial x^2. \quad (17)$$

This gives

$$\left(\frac{\partial^2}{\partial t^2} - \frac{\partial^2}{\partial x^2}\right)\psi' = \left(\tan^2(\varphi) - 1\right)\frac{\partial^2 \psi'}{\partial x^2}, \quad (18)$$

where $\tan^2(\varphi) = v^2$. After the integration over $x$ we get

$$\frac{\partial \psi'}{\partial x} = \frac{\lambda}{1-v^2}\bar{\chi}'^2 = 0. \quad (19)$$

Inserting this equation into Eq. (15) and taking into account that $\psi'^3 \cong 4\bar{\psi}'^3 \cos(\omega t')/3$ we find the following equation:

$$\left(\frac{\partial^2}{\partial t'^2} + \frac{\partial^2}{\partial x'^2}\right)\psi' + 4\psi' + 4\left(\mu - \frac{4\lambda^2/3}{1-v^2}\right)\psi'^3 = 0.$$

The solution of this equation is analogous to that of Eq. (8). In the laboratory reference frame this solution for the moving AC component of the vibrational soliton reads

$$\chi' = \sqrt{\frac{2}{3\mu - \frac{4\lambda^2}{1-v^2}}} \frac{\varepsilon \cos(\omega'(t+vx))}{\cosh(\varepsilon'(x-vt))}. \quad (20)$$

Analogously we get the moving DC component in the form

$$\psi' = \frac{2\lambda\varepsilon}{3\mu(1-v^2) - 4\lambda^2}\tanh(\varepsilon'(x-vt)). \quad (21)$$

There are two cases when the moving envelope wave can exist: a) with subsonic velocity $v<1$ and b) with supersonic velocity $v>1$. In the case a) the moving envelope wave exists only if its velocity is so low that the condition

$$|v|<\sqrt{1-4\lambda^2/3\mu} \tag{22}$$

is satisfied. In addition, the cubic anharmonicity should be sufficiently small. This motion has been observed in numerical simulations; see, e.g. [9-11]. The case b) corresponds to the supersonic motion in a chain with any cubic anharmonicity $\lambda$ but hard (positive) quartic anharmonicity $\mu>0$. As far as we know, this type of motion in chains with cubic and quartic anharmonicity has not been observed yet. For such motion the cubic anharmonicity plays the decisive role: the moving mode is stabilized by co-moving deformation of the lattice caused by this anharmonicity. The quartic anharmonicity may be absent or even soft ($\mu<0$). In this case the moving envelope wave exist if $v>\sqrt{1+4\lambda^2/3\mu}$. In this case the wave may have very large energy if its velocity gets close to $\sqrt{1+4\lambda^2/3\mu}$.

Note that the existence of presented here solution for supersonically moving large size vibrational solitons (envelope waves of self-localized of vibrations) with high frequency in monatomic chain indicates that analogous solitons may exist in other atomic chains, including polymers. A general argument in favor of this possibility is as follows: considered here equations of motion in the continuous limit of the chain may hold independently of the actual structure of the system under consideration in the atomic scale. Of special interest in this connection are proteins, being, in fact, biological polymers. Note that a different type of vibrational solitons – low-frequency collective motion in biomacromolecules – was considered by several authors; see, e.g. [16]. The possibility of fast motion of vibrational excitations with high frequency of vibrations in polymers may be important for understanding energy transport in these systems.

**Acknowledgement**.

This work was supported by institutional research funding (IUT2-27) of the Estonian Ministry of Education and Research.

**References**

1. A.M. Kosevich, A.S. Kovalev, Zh. Eksp. Teor. Fiz. **67**, 1793 (1974).
2. S. Dolgov, Sov. Phys. Solid State **28**, 907 (1986).
3. A.J. Sievers and S. Takeno, Phys. Rev. Lett. **61**, 970 (1988).
4. K.W. Sandusky, J.P. Page, and K.E. Schmidt, Phys. Rev. B **46**, 6161 (1992).
5. Y. S. Kivshar and N. Flytzanis, Phys. Rev. A **46**, 7972 (1992).
6. R.S. MacKay, S. Aubry, Nonlinearity **7**, 1623 (1994).
7. S. Flach, A. Gorbach, *Discrete breathers: advances in theory and applications*, Physics Reports **467**, 1-116 (2008).
8. M. Haas, V.Hizhnyakov, A. Shelkan, M. Klopov, A.J. Sievers, Phys. Rev. B **84**, 144303 (2011).
9. K. Hori, S. Takeno, J. Phys. Soc. Jpn. **61**, 2186 (1992).
10. S. R. Bickham, A. J. Sievers, and S. Takeno, Phys. Rev. B **45**, 10344 (1992).
11. S. A. Kiselev, S. R. Bickham, and A. J. Sievers, Phys. Rev. B **50**, 9135 (1994).
12. V. Hizhnyakov, M. Haas, A. Shelkan, M. Klopov, in: J.F.R. Archilla, N. Jimenez, V.J. Sanchez-Morsillo, L.M. Garcia-Raffi (eds) *Quodons in mica: nonlinear localized travelling excitations in crystals*, Springer (2015).
13. S. Flach , K. Kladko, Physica D **127**, 61 (1999).
14. S.R. Bickham, S.A. Kisilev, A.J. Sievers, Phys. Rev. B **47,** 14206 (1993).
15. D. Chen, S. Aubry, G.P. Tsironis, *Breather mobility in discrete $\phi^4$ lattices*, Phys. Rev. Lett. **77,** 4776 (1996). Nesterenko V.F., Prikl. Mekh. Tekh. Fiz. **5,** 136 (1983).
16. Z. Sinkala, Journal of Theoretical Biology **241** (4): 919–27 (2006). doi:10.1016/j.jtbi.2006.01.028.